\documentclass[runningheads]{llncs}
\usepackage[T1]{fontenc}
\usepackage{graphicx}
\usepackage{xcolor}
\usepackage{subfigure}%
\usepackage{adjustbox}

\begin{document}
%
%\title{Recent advances on Multiple Sclerosis Lesion Segmentation}
%\title{ICPR 2024 Competition on Multiple Sclerosis Lesion Segmentation (MSLesSeg) - Competition Report}
%oppure
\title{ICPR 2024 Competition on Multiple Sclerosis Lesion Segmentation - Methods and Results}
\titlerunning{MSLesSeg Competition}
\author{Alessia Rondinella\inst{1}\orcidID{0000-0002-6825-8708} \and
Francesco Guarnera\inst{1}\orcidID{0000-0002-7703-3367} \and
Elena Crispino\inst{2}\orcidID{0000-0001-9289-4926} \and
Giulia Russo\inst{3}\orcidID{0000-0001-6616-7856} \and
Clara Di Lorenzo\inst{4} \and
Davide Maimone\inst{5}\orcidID{0000-0003-1906-7953} \and
Francesco Pappalardo\inst{3}\orcidID{0000-0003-1668-3320} \and
Sebastiano Battiato\inst{1}\orcidID{0000-0001-6127-2470}}
\authorrunning{A. Rondinella et al.}
\institute{Department of Mathematics and Computer Science, University of Catania, Catania, Italy \and
Department of Biomedical and Biotechnological Sciences, University of Catania, Catania, Italy \and
Department of Drug and Health Sciences, University of Catania, Catania, Italy \and
UOC Radiologia, ARNAS Garibaldi, Catania, Italy \and
Centro Sclerosi Multipla, UOC Neurologia con Stroke Unit, Azienda Ospedaliera per l’Emergenza Cannizzaro, Catania, Italy}
\maketitle              % typeset the header of the contribution
\begin{abstract}
This report summarizes the outcomes of the ICPR 2024 Competition on Multiple Sclerosis Lesion Segmentation (MSLesSeg). The competition aimed to develop methods capable of automatically segmenting multiple sclerosis lesions in MRI scans. Participants were provided with a novel annotated dataset comprising a heterogeneous cohort of MS patients, featuring both baseline and follow-up MRI scans acquired at different hospitals.
MSLesSeg focuses on developing algorithms that can independently segment multiple sclerosis lesions of an unexamined cohort of patients. This segmentation approach aims to overcome current benchmarks by eliminating user interaction and ensuring robust lesion detection at different timepoints, encouraging innovation and promoting methodological advances.

%Multiple Sclerosis is a neurological disease that affects millions of people worldwide. In MS, different area of the brain become inflamed, damaging the myelin, the fatty tissue that surrounds and insulate nerve fibers. Magnetic Resonance Imaging is a fundamental tool to reach the diagnosis of MS and to monitor its progression and response to treatments. Accurate segmentation of MS lesions is essential for volumetric quantification of MS lesion load. However, manual selection of MS lesions on MRI scans is a very strenuous and time-consuming task and sufficiently large data sets with accurate manual segmentation by experts are still lacking. Consequently, the development of methods capable of automatically segmenting MS lesions is an unmet need and would represent a key step in advancing clinical management and optimizing treatment for people with MS. In conjunction with the ICPR 2024 conference, we propose a new lesion segmentation competition focused on improving the accuracy of MS lesion segmentation in MRI. We plan to provide participants with an extensively annotated dataset derived from a heterogeneous cohort of MS patients, which contains both baseline and follow-up MRI scan of each patient, acquired at different hospital. MSLesSeg focuses on developing algorithms that can independently segment MS lesions of an unexamined cohort of patients. This segmentation approach aims to overcome current benchmarks by eliminating user interaction and ensuring robust lesion detection at different timepoints, encouraging innovation and promoting methodological advances.

\end{abstract}
\section{Introduction}
\label{sec:introduction}
Multiple Sclerosis (MS) is a chronic neurological disease affecting millions of people worldwide. MS, causes inflammation in various areas of the brain, resulting in damage to the myelin, the fatty tissue that surrounds and insulates nerve fibers, which is crucial for proper neurological function \cite{lassmann2018multiple}. Magnetic Resonance Imaging (MRI) is a fundamental tool for diagnosing MS, monitoring its progression, and evaluating responses to treatments \cite{filippi2018}. Accurate segmentation of MS lesions is essential for volumetric quantification of lesion load, which helps in clinical decision-making and patient management. However, manual segmentation of MS lesions on MRI scans is a labor-intensive and time-consuming process that requires significant expertise. Moreover, large datasets with manual segmentations carried out by experts are limited, posing a significant challenge for the development and validation of automated segmentation methods \cite{molyneux1999visual}. This highlights the urgent need for developing automated methods capable of segmenting MS lesions, which would significantly enhance clinical management and treatment optimization for MS patients \cite{shoeibi2021applications}.
Recent advances in Deep Learning (DL) have shown significant potential in automating this task, leading to more consistent and efficient lesion segmentation \cite{rondinella2023boosting,rondinella2023enhancing}. As known, Deep learning methods require the availability of huge amount of data, therefore the lack of large labeled datasets represents a limitation to these studies. The scarcity of large, annotated datasets restricts the training and validation of deep learning models, potentially impacting their generalizability and performance on unseen data.
Our initiative aims to fill existing gaps in fully automated MS lesion segmentation by providing participants with an extensively annotated MRI dataset derived from a heterogeneous cohort of MS patients. A notable strength of our initiative lies in the substantial number of patients scans, surpassing publicly available datasets used for MS lesion segmentation. Furthermore, our dataset is particularly  advantageous because it is representative of real-world scenarios, with authentic MS patients; in fact, it is acquired "in daily practice", reflecting a heterogeneous and unconstrained acquisition environment. Table \ref{tab:dataset_comparison} presents a numerical comparison of the main datasets used for MS lesion segmentation, highlighting the advantages of the MSLesSeg dataset in relation to existing datasets.

\begin{table} [t]
\centering
\caption{Comparison between the proposed dataset and the main dataset used in MS Lesion segmentation. \textit{Dataset} indicates the name of the dataset, \textit{N. patients} denotes the number of patients in each dataset. \textit{M/F} denotes the Male/Female ratio. \textit{N. timepoints} is the number of timepoints. \textit{Age Mean (SD)} is the mean age (and standard deviation) in years, at baseline. \textit{Training cases} and \textit{Testing cases} are the division of data into training and testing sets.}\label{tab1}
\begin{adjustbox}{width={\textwidth}}%
\begin{tabular}{|l|l|l|l|l|l|l|l|}
\hline
Dataset & N. & M/F & N. & Age  & Follow-Up & Training   & Testing \\
 & patients & & timepoints & Mean (SD) & Mean (SD) & cases & cases \\
\hline
ISBI2015 & 19 & 4/15 & 2-6 & 43.5 (±10.3) & 1.0 (±0.18) & 5 & 14\\
MSSEG-2016 & 53 & 15/38 & 1 & 45.4 (±10.3) & N.A. & 15 & 38 \\
%MSSEG-2 & 100 & 2 & New MS Lesion & 40 & 60\\
\textbf{MSLesSeg} & \textbf{75} & \textbf{27/48} & \textbf{1-4} & \textbf{37 (±10.3)} & \textbf{1.27 (±0.62)} & \textbf{53} & \textbf{22}\\
\hline
\end{tabular}
\label{tab:dataset_comparison}
\end{adjustbox}
\end{table}

The goal of the MSLesSeg competition is to promote the development of algorithms capable of autonomously segmenting MS lesions in unseen MRI data series (T1-w, T2-w, and FLAIR). By ensuring robust lesion detection across different timepoints, this competition aims to push the boundaries of current benchmarks, encouraging innovation and promoting methodological advancements in the field of MS lesion segmentation.
We are confident that the MSLesSeg competition represents a significant step forward in the pursuit of automated MS lesion segmentation. By providing a diverse and comprehensive dataset, we aim to inspire the development of advanced algorithms that can significantly improve the clinical management of MS and optimize treatment strategies for patients worldwide.

The remainder of this paper is structured as follows: Section \ref{sec:dataset_and_metrics} introduces the dataset used in the competition, detailing the preprocessing steps and the lesion annotation process. Section \ref{sec:competition_overview} provides an overview of the competition, including its description, protocol, duration, and the metrics used for evaluating the results. Section \ref{sec:participants_methods} contains information about the participating teams and descriptions of the methods they proposed. Section \ref{sec:competition_results} presents the results achieved by the participants, along with the final leaderboard. Finally, Section \ref{sec:conclusion} concludes the paper with a discussion of the competition's outcomes and future directions.

\section{Dataset}
\label{sec:dataset_and_metrics}

\subsection{MSLesSeg Dataset}
For this competition, we provided participants with the MSLesSeg Dataset, a comprehensively annotated dataset containing labeled MRI scans acquired from various MRI scanners. The dataset includes MRI series from 75 patients aged between 18 and 59 years, with an average age of 37 (±10.3) years. Of these patients, 48 are women and 27 are men. The MRI scans were acquired at multiple timepoints, ranging from 1 to 4 per patient. Specifically, 50 patients had 1 timepoints, 15 patients had 2 timepoints, 5 patients had 3 timepoints and 5 patients had 4 timepoints. The interval between consecutive timepoints is approximately 1.27 (±0.62) years. In total, the dataset comprises 115 MRI data series. Each timepoint includes three different scan modalities: T1-weighted (T1-w), T2-weighted (T2-w), and Fluid-Attenuated Inversion Recovery (FLAIR). The dataset was meticulously preprocessed and annotated by experts, with lesion annotations performed on the FLAIR sequences and complementary T1-w and T2-w sequences used for comprehensive lesion characterization. The dataset was divided into training and test sets, with 53 scans allocated to the training set and 22 scans reserved for the test set.
The study was approved by the corresponding Hospital Ethics Committee and all patients gave their informed consent.

\subsection{Dataset preprocessing}
\begin{figure} [t]
\includegraphics[width=\textwidth]{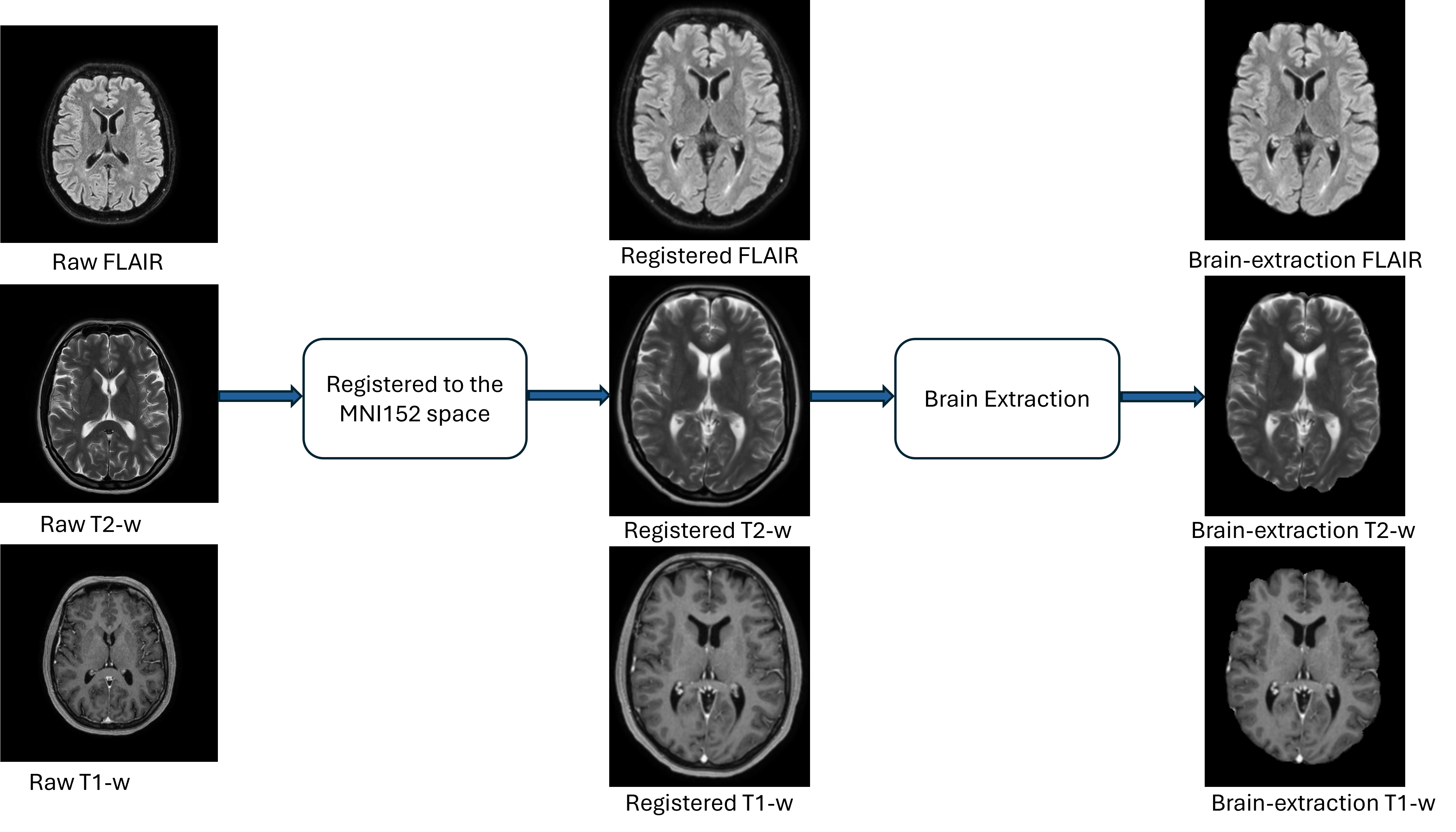}
\caption{Preprocessing steps: all brain MRI (T1-w, T2-w and FLAIR) were aligned to the standard $1 mm^3$ MNI space, then brain tissue was extracted.} \label{fig:preprocessing}
\end{figure}

\begin{figure}[t]
        \centering
        \subfigure[Flair]{\includegraphics[width=0.22\textwidth]{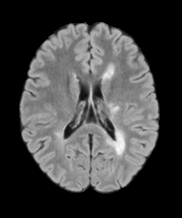} \label{subfig:flair}}
        \subfigure[T1-W]{\includegraphics[width=0.22\textwidth]{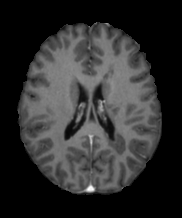} \label{subfig:t1}}
        \subfigure[T2-W]{\includegraphics[width=0.22\textwidth]{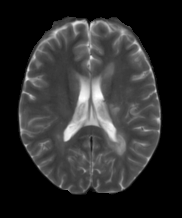} \label{subfig:t2}}

        \caption{Sample axial MRI images of the brain of an MS patient of the MSLesSeg Dataset in each modality of acquisition \ref{subfig:flair} FLAIR, \ref{subfig:t1} T1-weighted, \ref{subfig:t2} T2-weighted} \label{fig:mri}
\end{figure}

All MRI scans in the dataset underwent a series of preprocessing steps to ensure uniformity. At first, all scans were fully anonymized to protect patient privacy and then the scans were converted from the DICOM format to the NIFTI format, which is widely used in neuroimaging due to its compatibility and ease of use.
Subsequently, each MRI modality was co-registered to the $1 mm^3$ MNI152 isotropic template using FMRIB's Linear Image Registration Tool (FLIRT) \cite{jenkinson2002improved}, a fully automated instrument for brain image registration, ensuring that all scans are aligned to a common reference space.
After registration, brain extraction was performed using the Brain Extraction Tool (BET) \cite{smith2002fast}. BET effectively removes non-brain tissues, such as the skull and scalp, isolating the brain for subsequent analysis. This preprocessing pipeline ensures that all MRI scans are standardized, facilitating the development and evaluation of MS lesion segmentation algorithms.
The preprocessing pipeline is depicted in Figure \ref{fig:preprocessing}, while a proper example of images after preprocessing is shown in Figure \ref{fig:mri}.

\subsection{Lesion delineation}

\begin{figure}[t]
        \centering
        \subfigure[FLAIR]{\includegraphics[width=0.7\textwidth]{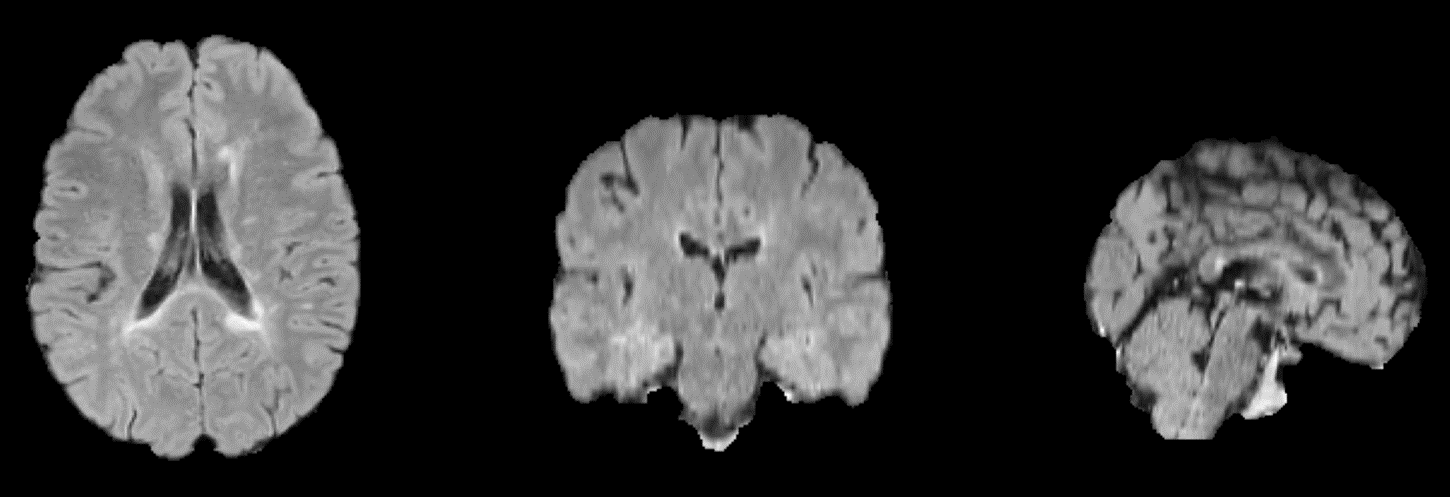} \label{subfig:flair_8}}
        \subfigure[Mask]{\includegraphics[width=0.7\textwidth]{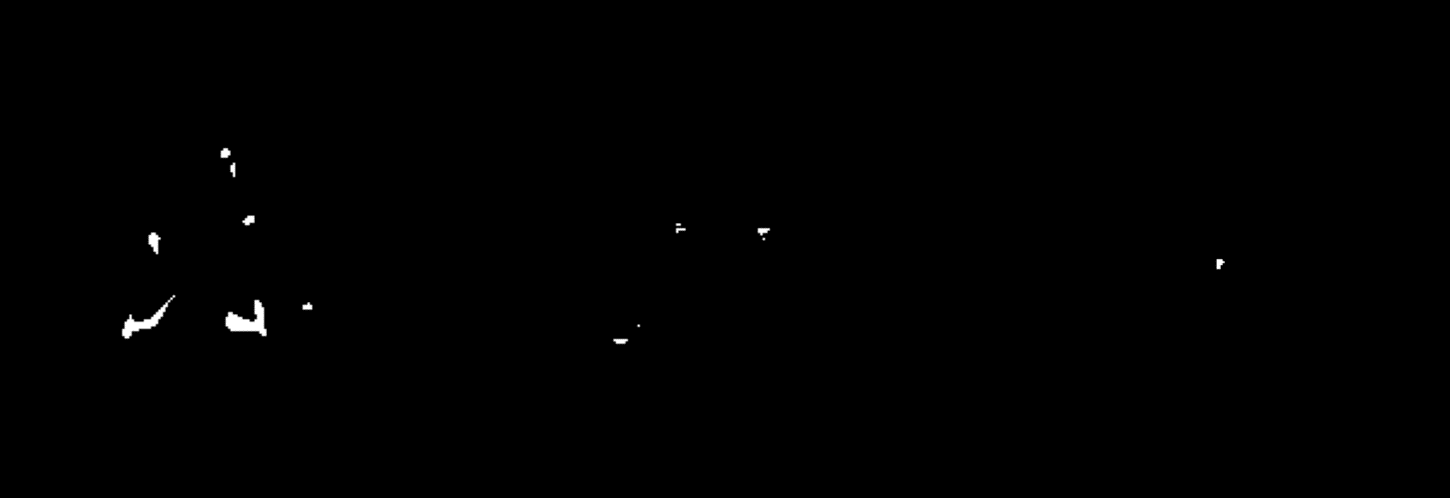} \label{subfig:mask_8}}

        \caption{Example of a (b) segmentation mask generated from the (a) FLAIR sequence. The images are presented in axial, coronal, and sagittal views.} 
        \label{fig:mask}
\end{figure}

The dataset was manually annotated to obtain the ground-truth masks of hyperintense lesions on FLAIR for each timepoint of every patient. Manual segmentation of MS lesions on MRI is inherently challenging due to the complexity of the task and the inherent variability among different expert annotators. 
To achieve a reliable ground-truth, the task was carried out by three experts. Lesion segmentation was performed primarily on the FLAIR modality, with comprehensive cross-checks on T2-w and T1-w modalities. The manual segmentation was conducted by a junior rater, who was trained by two senior experts: a senior neuroradiologist and a senior neurologist, both with extensive experience in MS.
Several training sessions were held between the junior rater and the two senior experts to establish a consistent lesion segmentation strategy and to introduce the junior rater with the segmentation tool, JIM9 \footnote[1]{https://www.xinapse.com/j-im-9-software/}. JIM9 is a sophisticated medical image processing software known for its advanced capabilities in image registration, segmentation, and analysis.
After the training sessions, the junior rater began labeling the 115 MRI data series. Throughout the process, additional meetings were held with the senior experts to validate the annotated scans. This iterative approach ensured that the annotations were accurate and consistent.
The manual segmentation protocol required independent segmentation for each patient and each timepoint to avoid bias. Segmentation was performed on the FLAIR images registered to the $1 mm^3$ isotropic MNI template, with T2-w and T1-w images used to confirm the presence of ambiguous or challenging lesions. After validation by the two senior experts, the resulting mask was deemed as ground-truth.
An example of a ground-truth mask is illustrated in Figure \ref{fig:mask}.

\section{Competition Overview}
\label{sec:competition_overview}

In this competition, the objective is the automatic segmentation of MS lesions. To tackle this task, we provided participants with MRI scans in three modalities: FLAIR, T1-w, and T2-w, along with their corresponding segmentation masks, as a ground-truth. The segmentation masks are binary, where white pixels represent the regions of the image corresponding to MS lesions, and black pixels represent the background.
Participants were asked to use any or all of these modalities, as well as the provided ground-truth, to develop deep learning-based algorithms capable of automatically generating MS lesion segmentation masks. MS lesions are represented as clusters of pixels in the ground-truth masks and can vary greatly in size and shape. These lesions are often very challenging to detect with the naked eye on MRI scans and require expert interpretation.
The goal is to develop deep learning algorithms that can segment MS lesions in a fully automated manner. An example of a segmentation mask is shown in Figure \ref{subfig:mask_8}.

\subsection{Competition Protocol and Duration}
During the competition, a dedicated website \footnote{https://iplab.dmi.unict.it/mfs/ms-les-seg/} was used to host all the rules and details useful for participants. This website will remain active even after the competition concludes, serving as a resource for participants and the wider research community.
The competition, which spanned three months, began with a team registration phase. Participants, after registering, were required to sign a Data Licence Agreement (DLA) before obtaining the training set; after DLA, the list of participants was published on the competition website.
After releasing the training set, participants were invited to develop methods for the automatic segmentation of MS lesions. Only fully automated methods were allowed. To ensure fair comparison among the participants, the data used during the training phase was limited to the dataset provided by the competition; no additional data was permitted. Afterwards, the test dataset was released, which included only the FLAIR, T1-w, and T2-w scans for each patient, but not the ground-truth masks. 
Alongside the release of the test dataset, we provided an evaluation script. This script allowed participants to calculate their scores using their methods on the training set scans by using a subset as a validation set, as they did not have access to the ground-truth of the test set scans. This same script was used by the organizers to generate the final leaderboard.
During the competition, each team was allowed to upload up to three submissions, with only the last submission being considered for the final results. 
Participants then moved to the submission phase, where they were required to submit a repository containing the segmentation masks generated by their proposed solutions. Additionally, each team was asked to produce a 1-2 page summary detailing their technical solutions to the competition.
After the submission deadline, scores were calculated only for participants who uploaded both the summary and at least one result file. All results were publicly displayed on the competition website, where the final leaderboard was also published at the end of the competition.

\subsection{Evaluation metrics}
The participants' methods were evaluated based on their accuracy in segmenting MS lesions, measuring the overlap between the predicted lesions by the proposed method and the actual lesions in the ground-truth mask. We used a metric based on overlap, specifically the Mean Dice Score (DSC) \cite{dice1945measures}, to determine the participants' ranking in the final leaderboard.
The evaluation of the models was conducted by comparing the submitted predicted segmentation masks of each teams with the ground-truth maskss. The DSC was the metric used, defined as follows:

\begin{equation}\label{eq}
DSC = \frac{2TP}{2TP+FP+FN}
\end{equation}

where TP, FP, and FN represent the number of true positive, false positive, and false negative pixels, respectively. The DSC is a widely used metric for measuring the similarity between two samples, especially in medical image segmentation tasks such as lesion segmentation, where precise delineation of objects is crucial. The DSC ranges from 0 to 1, where 0 indicates no overlap between the predicted and ground-truth masks, and 1 indicates a perfect overlap.
Specifically, we calculated the DSC for each data series in the test set using the formula (\ref{eq}). The final performance ranking of a participant was determined by the overall mean DSC  across all test set data series.

\section{Participants and Methods}
\label{sec:participants_methods}
A total of 28 teams registered for the competition, with 15 teams submitting their predictions for the final phase. Some teams were unable to submit their predictions due to the strict time constraints. Another possible reason could be the challenging nature of the dataset, which might have discouraged some participants.
Overall, we are very pleased with the contributions of the participants and hope that all participants, including those who were unable to submit their results on time, will continue working with the dataset to improve their segmentation methods. Below, we briefly describe the methods proposed by each team:\\
\textbf{MSUniCaTeam.}
The team, composed of Andrea Loddo, Alessandro Pani, Lorenzo Putzu, Luca Zedda, and Cecilia Di Ruberto, all from the University of Cagliari, proposed an ensemble method using Swin-UNETR \cite{hatamizadeh2021swin}, UNETR \cite{hatamizadeh2022unetr}, and SegResNet \cite{myronenko20193d} models. They preprocessed the data by merging contrasts, random cropping, and normalizing intensity. Each model was trained with the same number of epochs and batch size. They used Dice Loss for model training. The final predictions were produced by combining the models’ outputs through a weighted average based on the DSC from the validation set, enhancing the robustness of the final segmentation results.\\
\textbf{LST.}
The team, composed of Domen Preloznik and Ziga Spiclin from the Laboratory of Imaging Technologies, University of Ljubljana, further preprocessed the dataset using ANTsPy \footnote[2]{https://github.com/ANTsX/ANTsPy} to register T1-w images to the MNI ICBM 2009c atlas and computed mean transformations. They transformed all modalities and masks using this mean transformation, cropped images, and rescaled them. They trained their model using the nnU-Net \cite{isensee2021nnu} framework with custom configurations, experimenting with various initialization and training strategies. For evaluation, they used lesion-specific metrics from the Lesion-metrics repository \footnote[3]{https://github.com/Pubec/lesion-metrics} and generated segmentation masks using the best-performing model, applying inverse transformations to return images to their native space.\\
\textbf{UBHM.}
This team, consisting of Dana Dascalaescu from the University of Bucharest, Iuliana Georgescu from Helmholtz Munich, and Radu Tudor Ionescu from the University of Bucharest, employed the UNet++ architecture \cite{zhou2018unet++} to detect MS lesions, experimenting with different backbones. They used a combination of binary cross-entropy and Dice loss functions for training. In their two-stage approach, they first trained a classifier to distinguish slices with lesions and then applied a UNet++ model only to those slices. They also experimented with a patch-level approach, dividing slices into 2x2 parts and training separate models for each part. Finally, they created ensembles of the different models using a soft voting scheme to boost performance.\\
\textbf{UGIVIA-UIB.}
This team, composed of Aina Maria Tur Serrano, Gabriel Moyà Alcover, and Francisco José Perales López, all from the University of the Balearic Islands, focused on using only the FLAIR sequences, resizing 2D slices through cropping and padding. They employed a modified 2D U-Net architecture \cite{ronneberger2015u}, adjusting it to handle the resized input. The model consists of an encoder-decoder structure with a total of 19 convolutional layers. Data augmentation techniques, such as rotation, scaling, and shearing, were applied to increase the training set by generating 4000 new images. The final segmentation masks were produced by resizing the predictions back to their original size.\\
\textbf{MadSeg.}
The team, composed of Liang Shang, Zhengyang Lou, William Sethares, Vivek Prabhakaran, Veena Nair, Nagesh Adluru, all from the University of Wisconsin-Madison, employed two labeling strategies to enhance model training: Multi-Size Labeling (MSL) and Distance-Based Labeling (DBL). MSL classifies lesions into small, medium, and large categories based on their volume, while DBL labels lesion voxels according to their distance from non-lesion areas. They trained models using 5-fold cross-validation with various training schemes: the default scheme with nnU-Net \cite{isensee2021nnu} and a combination of Dice+CrossEntropy loss, a focal scheme with Dice+Focal \cite{lin2017focal} loss, an MSCSA scheme integrating the attention-based MSCSA module \cite{shang2023vision}, and two variants with Res UNet and Mamba-based LightM-UNet \cite{liao2024lightm}. For postprocessing, they created ensembles of predictions for each labeling strategy, combining results and applying linear interpolation for small lesions and relying on the DBL ensemble for large lesions. Finally, they filtered out small lesions based on a probability threshold to obtain the final prediction results.\\
\textbf{BrainS.}
The team, composed of Edoardo Coppola, Mattia Savardi, Alberto Signoroni, and Sergio Benini, all from the University of Brescia, used the nnU-Net self-configuring framework \cite{isensee2021nnu} for segmentation. Initially, they conducted a 5-fold cross-validation using only FLAIR images, followed by another 5-fold cross-validation using T1-w, T2-w, and FLAIR images. The multi-modality approach showed improved performance wrt single-modality approach. After confirming the benefit of using multiple modalities, they performed a statistical analysis on the training and test sets to check for covariate shifts, which indicated no significant distribution shift. Finally, they involved an experienced neuroradiologist to evaluate the model results and selected the best-performing model with a good precision-recall balance for the final evaluation.\\
\textbf{BeckLab.}
The team, composed of Francesco La Rosa, Emma Dereskewicz, and Erin S. Beck, all from the Icahn School of Medicine at Mount Sinai, trained two nnU-Net \cite{isensee2021nnu} 3D models with a 5-fold cross-validation on the training dataset, using the default hyperparameter configuration. The first model used only FLAIR images, while the second model used T1w, T2w, and FLAIR images. After visual inspection of the validation masks, they found that combining the outputs from both models enhanced MS lesion detection and improved overall lesion segmentation. This combined mask was used as the final output of their method.\\
\textbf{M3S: MeetMIALMS.}
The team, composed of Pedro M. Gordaliza from the CIBM Center for Biomedical Imaging, University of Lausanne and Lausanne University Hospital, Federico Spagnolo from the Translational Imaging in Neurology, Maxence Wynen from the Catholic University of Louvain, Nataliia Molchanova from the University of Lausanne and Lausanne University Hospital, University of Applied Sciences of Western Switzerland, CIBM Center for Biomedical Imaging, and Meritxell Bach Cuadra from the CIBM Center for Biomedical Imaging, University of Lausanne and Lausanne University Hospital, employed the nnU-Net framework \cite{isensee2021nnu}, utilizing both U-Net \cite{ronneberger2015u} and the U-Mamba \cite{ma2024u} architectures. The nnU-Net automatically configures network architecture, training, and testing pipelines for a given dataset, featuring a five-level 3D CNN, extensive data augmentation, and model ensembling to enhance generalization. The U-Mamba integrates CNN and State Space Sequence Models (SSMs) to combine local feature extraction with long-range dependency capture. They explored three models: baseline nnU-Net, nnUNetUMambaBot (with the U-Mamba blocks at the network bottleneck), and nnUNetUMambaEnc (with the U-Mamba blocks throughout the entire architecture), using FLAIR images alone and in combination with T1-w and T2-w images. T1-w and T2-w images were registered to FLAIR images using ANTS. Six models were trained, and the ensemble achieving the highest DSC on a patient-wise basis was selected. The nnUNet-FLAIR model demonstrated the best performance with a DSC of 0.773 and was used for the test dataset.\\
\textbf{Student.}
The team, composed of Kwaku Agyapong from the Université C\^ote d'Azur, used T2-w and FLAIR modalities from the MRI scans, extracting 2D axial views for training and validation. A U-Net architecture with EfficientNet-b0 as the backbone was employed. They used weighted binary cross-entropy loss and the DSC for evaluation. Data augmentation and callbacks like learning rate reduction and early stopping were implemented to improve performance. \\
\textbf{MMS.}
The team, composed of Aswathi Varma, Benedikt Wiestler, and Daniel Scholz, all from the TU Munich, utilized FLAIR and T2-w images for their segmentation approach, leveraging the nnU-Net framework \cite{isensee2021nnu} enhanced by a novel augmentation technique called Global Intensity Non-linear (GIN) augmentations. GIN involves using randomly initialized convolution layers \cite{choi2023progressive} and non-linear functions like ReLU to generate diverse appearances for training images while preserving anatomical structures. This method helps the network focus on shape invariant information. For each training image, ten GIN-augmented views were created and used along with the original images to train the nnU-Net segmentation network, employing a 5-fold cross-validation with a 3D full-resolution configuration.\\
\textbf{BronTeam.}
The team, composed of Lehel D\'enes-Fazakas, Barbara Simon, Adam Hartveg, L\'aszl\'o Szil\'agyi, all from Obuda University, adapted a U-Net network for the competition, consisting of four encoder and four decoder blocks with a bridge in between. Each block performed 2+1 dimensional convolutions, using 3D convolution operations. Training involved the Adam optimizer, using a composite loss function combining Categorical Cross Entropy and Dice loss.\\
\textbf{INTELLIGO Labs.}
The team, composed of Federico Cunico, Giorgio Dolci, Ilaria Boscolo Galazzo, Gloria Menegaz, and Marco Cristani, all from the University of Verona, implemented a 3D U-Net model with attention gates on residual connections \cite{oktay2018attention} processing only the FLAIR scan. The architecture included 3 encoder layers, a bottleneck layer, and 3 decoder layers, expanding from a 64-dimensional space in the encoder to a 256-dimensional space in the bottleneck before decoding to the final output. The weights were initialized using the Kaiming normal distribution, and the model was trained using reduced precision (16-bit) to manage memory limitations, allowing a larger batch size. A grid search was conducted to find the optimal seed for weight initialization.\\
\textbf{LA2I2F.}
The team, composed of Lauro Snidaro, Federico Urli, Ehsan Rassekh and Michele Somero, all from the University of Udine, proposed a 3D U-Net model, known as 3dSeUnet, designed for volumetric data segmentation. The architecture includes an encoder with convolutional blocks and squeeze-and-excitation (ssCE3d) layers, a bottleneck for processing downsampled features, and a decoder with transposed convolutions and skip connections. The custom ssCE3d layer enhances feature recalibration using global average pooling and a multi-layer perceptron (MLP). Their custom BasnetLoss combines binary cross-entropy (BCE), structural similarity index (SSIM), and intersection over union (IoU) losses to guide training. They used the RMSprop optimizer with a learning rate of 0.0001, aiming for accurate and perceptually similar segmentations with high overlap with the ground truth.\\
\textbf{Golestan.}
The team, consisting of Majid Ziaratban from Golestan University, proposed a segmentation approach consisting of three main steps: preprocessing, segmentation, and post-processing. In preprocessing, each slice of an MRI from the training set was extracted and saved as a single image, generating a new dataset of over 126,000 images. Twenty combinations were created for each image with potential lesions. For segmentation, they used a 2D U-Net-based network. During post-processing, the segmentation results of the 20 combinations per slice were combined, and all segmented slices were assembled to produce the final mask.\\
\textbf{AdasLab.}
The team, composed of Marcos D\'iaz-Hurtado from the Universitat Oberta de Catalunya, Ferran Prados Carrasco from the Universitat Oberta de Catalunya, Jordi Casas Roma from the Universidad Aut\'onoma de Barcelona, Eloy Mart\'inez-Heras from the Neuroimmunology and Multiple Sclerosis Unit, Laboratory of Advanced Imaging in Neuroimmunological Diseases, Hospital Clinic Barcelona, Institut d’Investigacions Biomèdiques August Pi i Sunyer and Universitat de Barcelona, employed a method based on the nnU-Net architecture \cite{isensee2021nnu}. This approach handles both cross-sectional and longitudinal imaging data with only FLAIR images. They also employ a Monte Carlo strategy to generate synthetic lesion masks, simulating different lesion scenarios by applying changes like erosion, removal, and dilation. This method creates a robust training set, improving the model's generalization and reliability. 

%\paragraph{ICAI.} \textcolor{red} {non ha risposto alla mail}
%The team, composed of Esteban J. Palomo Ferrer, Iván García Aguilar, Ezequiel López Rubio, Rafael M. Luque Baena, all from the University of Malaga, implemented a segmentation process involving preprocessing steps such as image enhancement using the Real-ESRGAN model and resizing operations. They utilized a modified U-Net architecture with an adjusted loss function to improve MS lesion segmentation. The Real-ESRGAN model enhanced image quality by applying super-resolution, which provided more detailed information about the brain's anatomical structures. The images were then resized back to their original dimensions. The loss function was adjusted to address class imbalance and enhance segmentation performance by incorporating weighted loss and Dice loss.
\section{Competition Results}
\label{sec:competition_results}

\begin{table}
\caption{Final leaderboard, displaying the results obtained by each participating team in the MS lesion segmentation competition. The columns include the following information: the rank position of each team (column \textit{Ranks}), the team name (column \textit{Team Name}), the organization they represent (column \textit{Organization}), a summary of the methods they employed (column \textit{Methods}), the imaging modalities used (column \textit{Modalities}), and the mean DSC achieved on the test set (column \textit{Mean DSC}). This comprehensive overview highlights the diversity of approaches and the performance metrics achieved in the competition.}\label{tab2}
\centering
\begin{adjustbox}{width={\textwidth}}%
\begin{tabular}{|c|c|c|c|c|c|}
\hline
Rank & Team Name & Organization &  Methods & Modalities & Mean DSC\\
\hline
1 & MadSeg & University of & Ensembles of  & FLAIR, T2-w, &  0.7146\\
 & & Wisconsin-Madison & labeling strategies & T1-w & \\
 \hline
2 & BrainS & University of Brescia &  Statistical analysis & FLAIR, T2-w, & 0.7083\\
 & & & and nnU-Net & T1-w & \\
 \hline
3 & M3S:  & CIBM Center for & Modified nnU-Net & FLAIR, T2-w, & 0.7079\\
 & MeetMIALMS & Biomedical Imaging, & labeling strategies & T1-w & \\
 &  & Switzerland &  & & \\
\hline
4 & AdasLab & Universitat  & augmentation technique & FLAIR & 0.6974\\
 &  & Oberta de Catalunya & and nnU-Net & & \\
\hline
5 & MMS & TU Munich & augmentation technique & FLAIR, T2-w & 0.6859\\
 & & & and nnU-Net & & \\
\hline
6 & LST & Faculty of ,  & ANTsPy preprocessing, & FLAIR, T2-w, & 0.6783\\
 & & Electrical Engineering & nnU-Net & T1-w & \\
 & & University of Ljubljana &  & & \\
 \hline
 
7 & BeckLab & Icahn School of   & nnU-Net & FLAIR, T2-w, & 0.6754\\
 & & Medicine at Mount Sinai & & T1-w & \\
\hline
8 & MSUniCaTeam & University of Cagliari &  Ensemble of  & FLAIR, T2-w, & 0.6508\\
 & & & Swin-UNETR, UNETR, & T1-w & \\
& & & SegResNet & & \\
\hline
9 & Golestan & Golestan university &  2D U-Net & NA & 0.6503\\
\hline
10 & LA2I2F & University of Udine & 3D U-Net & FLAIR & 0.6446\\
 & &  & with custom loss & & \\
\hline
11 & UBHM & University of Bucharest, & UNet++, & FLAIR & 0.6357\\
 & & Helmholtz Munich  & ensembles of models &  & \\
\hline
12 & UGIVIA-UIB & University of  & 2D U-Net & FLAIR & 0.6101\\
 & Team & the Balearic Islands  &  & & \\
\hline
13 & BronTeam & John von Neumann & 3D U-Net & FLAIR, T2-w, & 0.5683\\
 & & Faculty of Informatics,  &  & T1-w & \\
  & &  Hungary  &  & & \\
\hline
14 & INTELLIGO  & University of Verona &  3D U-Net & FLAIR & 0.5471\\
& Labs & & with attention gate & & \\
\hline
15 & Student & University of  & EfficientNet-b0 & FLAIR, T2-w & 0.4985\\
 & & Cote d'Azur & & & \\
%\hline
%16 & ICAI & University of Malaga & Super-resolution & FLAIR, T2-w, & 0.2351\\
% & & & and modified U-Net & T1-w & \\

\hline
\end{tabular}
\end{adjustbox}
\end{table}

Table \ref{tab2} reports the final leaderboard. Each team was allowed to make up to three submissions, but the table displays only the result of the last submission for each team. Submissions are ranked according to the mean DSC metric. The online leaderboard can be viewed on the competition website.
Looking at the results in the table, we can see that the team \textit{MadSeg}, with all members from the University of Wisconsin-Madison, achieved the best performance. Their approach involved multi-size and distance-based labeling strategies for data augmentation. They trained several models, each with different loss functions and architectures. The predictions from these models were ensembled and post-processed to enhance segmentation results. This sophisticated strategy led them to attain a DSC of $0.7146$ on the test set, which is $0.0063$ points above \textit{BrainS}, the second-place team.
\textit{BrainS}, with all members from the University of Brescia, utilized a method based on the nnU-Net framework, performing 5-fold cross-validation with both single-modality (FLAIR) and multi-modality (T1w, T2w, FLAIR) approaches. Their method confirmed the benefits of multi-modality through statistical analysis, and they selected the best model based on neuroradiologist evaluation and precision-recall balance, achieving a DSC of $0.7083$.
The third-place team, \textit{M3S: MeetMIALMS}, with members from Catholic University of Louvain, Translational Imaging in Neurology, University of Lausanne and Lausanne University Hospital, University of Applied Sciences of Western Switzerland, and CIBM Center for Biomedical Imaging, used an ensemble of networks based on customized nnU-Net architectures, leveraging FLAIR modality alone and in combination with T1-w and T2-w modalities. They experimented with various model configurations and MRI sequences, ultimately selecting nnUNet-FLAIR, which achieved the highest DSC for their validation set. \textit{M3S: MeetMIALMS} secured a DSC of $0.7079$, just $0.0004$ points behind \textit{BrainS}, demonstrating the competitiveness of the top-performing teams. A closer examination of the results reveals that the most challenging lesions for segmentation were typically those with smaller sizes and those located near the ventricles. These lesions often have lower contrast against surrounding brain tissues, making them difficult to distinguish. Additionally, periventricular lesions, i.e. adjacent to the ventricles, are particularly challenging due to their proximity to areas with high-intensity signals such as the cerebrospinal fluid in FLAIR sequences, which can obscure lesion boundaries.
Another common source of segmentation errors involved hyperintensities in the white matter, visible in FLAIR images, which can be incorrectly identified as MS lesions. These hyperintensities, which may result from aging or other neurological conditions, frequently mimic the signal characteristics of MS lesions. The presence of these non-MS hyperintensities, especially in models that use only FLAIR images, poses a significant challenge, as they can lead to false-positive segmentations.
A key strength of the dataset used in this competition is the inclusion of T1-w and T2-w scans, which provide complementary information to confirm the presence and location of MS lesions. These additional modalities help distinguish true MS lesions from other high-intensity areas or non-lesional hyperintensities. This is reflected in the approaches of the three top-performing teams, all of which utilized all three available MRI modalities (FLAIR, T1-w, and T2-w) to achieve more accurate lesion segmentation. By leveraging the combined information from multiple sequences, these models were better able to mitigate the challenges posed by small lesions and periventricular regions, and differentiate true lesions from other hyperintensities.
Interestingly, models that incorporated more advanced techniques, such as synthetic data augmentation, ensemble learning, and innovative labeling methods, further enhanced their ability to differentiate true MS lesions from challenging cases. However, even with these techniques, the consistent segmentation of small lesions and lesions near high-intensity regions remained difficult, highlighting areas for potential future improvement. From an organizational standpoint, the competition proceeded smoothly without significant issues or unexpected difficulties. The submission process, which involved submitting a repository containing the segmentation masks generated by the proposed solutions on the test set, worked well. We plan to organize the competition again next year, but this time with a larger dataset and additional tasks for efficient MRI-based MS lesion analysis.

\section{Conclusion}
\label{sec:conclusion}
This paper summarizes the results of the successful submissions for the first competition on multiple sclerosis lesion segmentation. The outcomes of this competition demonstrate that deep learning algorithms can achieve promising results in MS lesion segmentation. However, these algorithms still need refinement to improve the detection of small and challenging lesions. For broader clinical adoption, model interpretability is crucial. Models that provide insights into their decision-making processes, such as via saliency maps or attention mechanisms, are more likely to be trusted in clinical environments. Future iterations of the MSLesSeg competition should emphasize interpretability alongside segmentation accuracy to facilitate the integration of these advanced models into clinical practice, ultimately enhancing patient care and treatment management. Furthermore, scalability is key for clinical implementation. The top-performing models, using techniques like ensemble learning and synthetic data augmentation, require significant computational resources. Nonetheless, ongoing advancements in hardware and optimization techniques can help mitigate these demands. Future research should focus on balancing accuracy and efficiency to ensure applicability across diverse clinical settings. In conclusion, this competition highlights new opportunities and challenges in the analysis of MS lesions from MRI scans. We anticipate that in the near future, deep learning algorithms will further demonstrate their potential to enhance the clinical management of MS and optimize treatment strategies for patients worldwide. Moreover, with the release of our public dataset, we expect a surge in research activity in this critical area, driving further advancements and innovations.

\subsubsection{Acknowledgements} 
Alessia Rondinella is a PhD candidate enrolled in the National PhD in Artificial Intelligence, XXXVII cycle, course on Health and life sciences, organized by Università Campus Bio-Medico di Roma. Francesco Guarnera is funded by the PNRR MUR project PE0000013-FAIR. 


\end{document}